# Observation of the evolution dynamics from starting to shutting of SWNT-mode-locked fiber laser

Yudong Cui, Xiankun Yao, Weiwang Hu, and Xueming Liu

State Key Laboratory of Modern Optical Instrumentation, College of Optical Science and Engineering, Zhejiang University, Hangzhou 310027, China

*Corresponding author: liuxueming72@yahoo.com

**Abstract**:

Dispersive Fourier transform (TS-DFT) technique opens a fascinating pathway to explore the ultrafast non-repetitive even, which has been employed to study the build-up process of mode-locked lasers. Here the whole evolution dynamics (from starting up to shutting down) of conventional soliton (CS), stretched pulse (SP) and dissipative soliton (DS) are investigated by using TS-DFT technique. The relaxation oscillation can be always observed before the formation of stable pulse operation, which is stemmed from the inherent advantage of the single-walled carbon nanotube. However, owing to the different pulse features, they exhibit the distinct evolution dynamics in the starting and shutting processes. Some critical phenomena are observed, including transient complex spectrum broadening and frequency-shift interaction of SP and picosecond pulses. These results could further deepen the understanding of the mode-locked fiber laser from the real-time point of view and is helpful for the laser design and applications.

**Introduction**

Mode-locked fiber lasers delivering ultrafast pulses have been employed in numerous applications as diverse as ophthalmology, micromachining, medical imaging and precision metrology [1]-[3], because fiber lasers maintain several inherent advantage, high electrical efficiency, lower maintenance, higher reliability, smaller footprint, and easier transportability [4][5]. So far, fiber laser are generally implemented with passively mode-locked techniques, such as nonlinear polarization rotation, nonlinear optical loop mirror, semiconductor saturable absorber mirror, single-walled carbon nanotube (SWNT) and two-dimensional semiconductor materials [6]-[8]. Amongst them, SWNTs are the promising candidate for the ultrashort pulse generation because they possess ultrafast excited state carrier dynamic and high optical nonlinearity [8][9].

To pursue the higher pulse energy, researchers successively proposed conventional soliton (CS), stretched pulse (SP, also named dispersion-managed soliton) and dissipative soliton (DS) by engineering the intracavity group velocity dispersion (GVD) [7][10]. The dynamics of mode-locked fiber laser and its output features depend on the dispersion map of laser oscillator [10]-[12]. Soliton fiber lasers are generally constructed with anomalous-dispersion condition, and pulses can be sustained through the balance of nonlinear and dispersive phase shifts [11]. So it is called CS to distinguish from other types of soliton in mode-locked fiber lasers [13]. When the net dispersion is near-zero, stretched pulse forms, which can decrease the intracavity nonlinearity by periodically stretching and compressing [11][13]. Under the net normal or all normal cavity dispersion, the pulse-shaping dynamics are dominated by the gain and loss with the assist of dispersion, nonlinearity and spectral filtering [7]. Dissipative soliton are always highly-chirped so that it can hold higher pulse energy and nonlinear phase shift [14].

The formation of soliton pulses is a process of phase locking of a large number of longitudinal cavity modes [15]. The studies about the starting dynamics of passive mode-locking lasers have helped researchers understand the build-up time, the capability of self-starting, and Q-switched instability, which is crucial to applications [16]-[18]. Real-time oscilloscope is the common tool to record the temporal evolution process, while the spectral information and pulses closed or even overlapped to each other cannot be resolved due to the limited electronic bandwidth [2]. Achieving a temporally and spectrally resolved study of the transient dynamics is always a challenge because every transient even possesses a singular occurrence with unique spectro-temporal features. Recent developed time-stretch dispersive Fourier transform (TS-DFT) technique provides a powerful way for real-time, single-shot measurements of ultrafast phenomena [19]. This technique helps scientists to experimentally resolve the evolution of femtosecond soliton molecules [20][21], the internal motion of dissipative optical soliton molecules [22], and the dynamics of soliton explosions [23]. TS-DFT technique was also employed to measure the build-up process of soliton [2][21][24]-[27]. These works mainly focus on the starting dynamics of soltion [2][25], soliton molecule [21][27], and multi-pulse operation [28] in the mode-locked fiber laser under anomalous dispersion. The spectral build-up process of dissipative soliton in the net-normal dispersion regime was reported in the recent work [26].

However, the starting dynamics of stretched pulse has not been studied experimentally no matter in temporal or spectral domain. Additionally, the evolution when the mode-locked fiber lasers are shut down also remains unclear. Circulating pulses can be amplified for thousands of times after the pump power is switched off, as the roundtrip time is far less than the relaxation time of the gain fiber. Moreover, the switching time is generally more than several milliseconds. Considering the complicated operation condition of fiber laser and distinct pulse characteristics, some interesting ultrafast phenomena may happen in the shutting process of mode-locked laser.

In this work, the whole evolution dynamics (from starting up to shutting down) of a SWNT-mode-locked fiber laser are measured via TS-DFT technique. By managing the intra-cavity net dispersion, CS, SP and DS are achieved under negative, near-zero and positive dispersion condition in the mode-locked laser, respectively. Their starting and shutting processes can be obtained with the high-speed photodetectors and oscilloscope. The build-up processes of CS, SP and DS both contain the relaxation oscillation, the Q-switched lasing and stable pulse operation stages, but display the distinct transient evolution dynamics. And CS, SP and DS also experience the different dynamics and energy fluctuation when the laser is shut down. Several new physical evolution processes are observed in the measurement and these results are helpful for the understanding of the mode-locked fiber laser.

**Experimental setup**

Figure 1 shows the schematic diagram of the nanotube-mode-locked fiber laser and the measurement scheme. The oscillator consists of a SWNT-based mode locker, a 7-m-long erbium-doped fiber (EDF) with ~4 dB/m absorption at 980 nm, a wavelength-division-multiplexed (WDM) coupler, an optical coupler, a polarization-independent isolator (PI-ISO) and a polarization controller (PC). A 980 nm laser diodes (LDs) provide pumps via a 980/1550 nm WDM coupler. An optical switcher is placed between LD and WDM to switch on or off the laser. A coupler with 10% output port is used to extract pulses from the cavity and PI-ISO ensures the unidirectional operation. PC is utilized to optimize the mode-locking performance by adjusting cavity linear birefringence. The fabrication procedure and parameters of the SWNT-based mode locker can be found in Ref. [29]. A section of standard single-mode fiber (SMF) is inserted into the cavity to adjust the net dispersion. The dispersion parameters of EDF and SMF are about −20 and 17 ps/nm/km at 1550 nm, respectively.

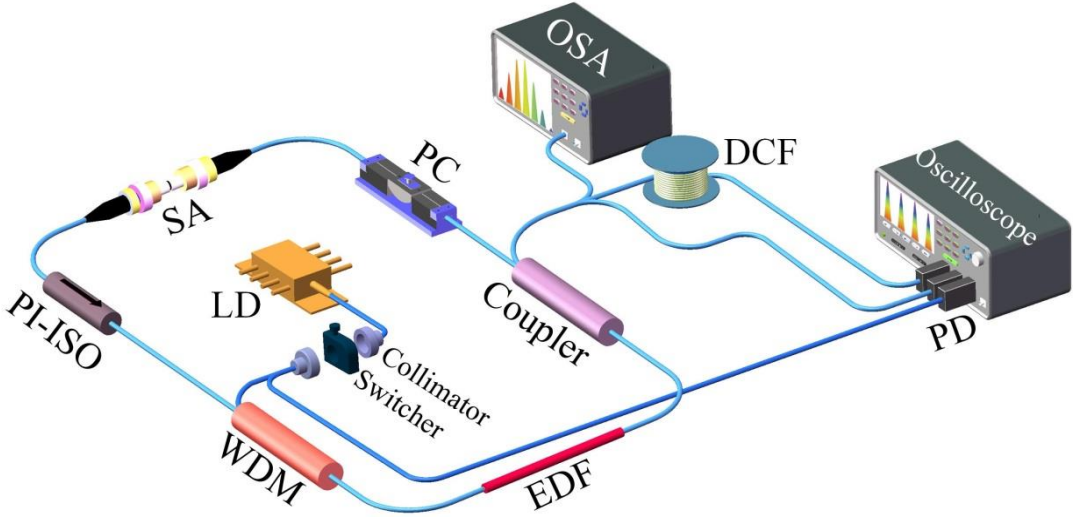

Fig. 1. Schematic diagram of the experimental setup. EDF, erbium-doped fiber; PC, polarization controller; LD, laser diode; SA, saturable absorber; WDM: wavelength-division-multiplexed coupler; PI-ISO: polarization-independent isolator. OSA: optical spectrum analyzer, DCF: dispersion compensation fiber.

Firstly, ~17-m SMF is in the cavity and the net dispersion is negative of ~-0.2 $ps^2$. In this case, CS with typical Kelly sidebands can be achieved [11]. Then SMF is gradually cut short carefully to adjust the cavity dispersion to zero. Here the length of SMF in the oscillator is ~ 10 m and the dispersion is ~-0.004$ps^2$. Finally, the inserted SMF is removed from the cavity and only the pigtail of devices is remained. The totally cavity length is ~14 m, and the intracavity net dispersion is ~0.02 $ps^2$, which can support the generation of DS. By managing the intra-cavity dispersion, CS, SP and DS are achieved in the nanotube-mode-locked fiber laser, respectively.

The real-time temporal detections for solitons are recorded with two high-speed photodetectors and a real-time oscilloscope. The real-time spectral information can be obtained by dispersing pulses in an ~5-km dispersion-compensating fiber (DCF) prior to detection. The spectral information could be mapped into the temporal waveform via DCF. The dispersion of DCF is about -160 ps/(nm km). The time-averaged spectra are measured via an optical spectrum analyzer. When the switcher is set as on or off, pump power can be transmitted or blocked. As a result, the fiber laser can start up or shut down. In order to monitor the starting and shutting dynamics of mode-locked fiber laser, a part of pump power is split and used as the triggering signal of oscilloscope, as shown in Fig. 1.

## Starting and shutting dynamics of mode-locked fiber laser

Figure 2 demonstrates the whole spectral evolution dynamics from starting up to shutting down for a SWNT-mode-locked fiber laser under distinct dispersion conditions. The starting and shutting processes of CS, SP and DS are shown in Fig. 2(a), (b) and (c), respectively. Benefitting from the SWNT SA, the start-up process in Fig. 2(a) is similar to our previous results, including the relaxation oscillation, Q-witched lasing stage and stable soliton operation [30]. While obvious different evolution profile of the intensity can be observed due to the different laser cavity parameters (e.g. the length of fiber, SA, the gain coefficient) and initial conditions (e.g. pump power). The Q-switched lasing here is much stronger than the relaxation oscillation, and the build-up time is about 1.8 millisecond that is longer than the former works [30]. When pump power is blocked, CS pulses could last ~0.3 ms continuously, and then vanish rapidly. That may be attributed to the mode locker whose loss would increase exponentially as the pulse energy decreases.

In the left part of Fig. 2(b), the relaxation oscillation can be also observed in the build-up process of SP. The formation of SPs costs ~1 ms, but, after that, a periodical intensity fluctuation last ~4.5 ms. The modulation should originate from the relaxation oscillation as the period at the beginning (from 2 ms to 3 ms) is identical to the that of relaxation oscillation. In fact, the relaxation oscillation can induce the Q-switched lasing in the build-up process of mode-locked laser [21]. However, it need more time to quiet down the Q-switched instability for SP. The shutting process for SP in the right side of Fig. 2(b) have the distinct evolution dynamics in comparison with CS (Fig. 2(a)). The intensity of SPs is modulated intensively before pulses vanish, that is the results of the common interaction of EDF and SA (this is also a kind of Q-switched instability). Owing to the larger pulse energy and the narrower pulse width, SPs can be re-amplified many times in the shutting process. It should be noted that the evolution dynamics would be influenced by many other factors, e.g., the pump intensity, turn-off time and SA. Here, only one SWNT SA is used in the experiment, and the pump intensity for SP (~30 mW) is slightly larger than CS (~22 mW). The turn-off time can be set from hundreds of nanoseconds to several milliseconds. With longer turn-off time, CSs generally vanish quickly firstly and SP experience longer Q-switched instability.

For DS, the evolution in the starting process seems to be similar with CS except that one lasing spike is much stronger, as shown in the left side of Fig. 2(c). It may result from the shorter cavity length and the higher pump power. The characteristics of relaxation oscillation is determined by the cavity structure and parameters [31]. The shutting process of DS in the right side of Fig. 2(c) exhibits obvious intensity fluctuation but the different evolution intensity envelop from SPs in Fig. 2(b), because of the distinct pulse energies, widths and evolution traces.

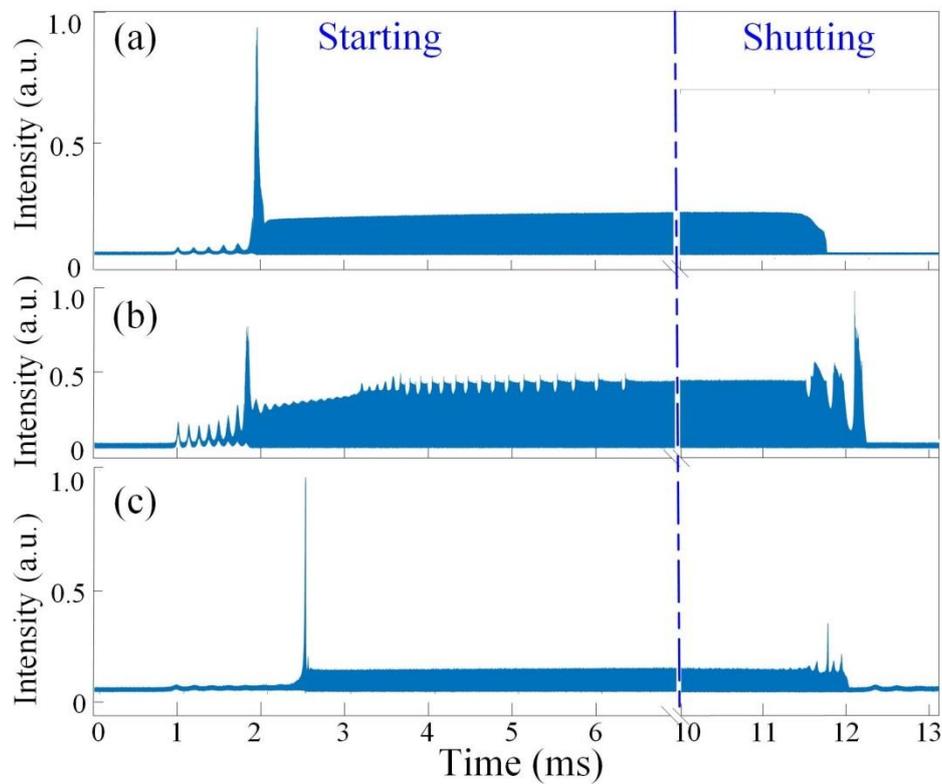

Fig. 2. Real-time observation of starting (Left) and shutting dynamics (Right) for (a) CS, (b) SP, (c) DS. These are the direct detection results of oscilloscope after the transmission in DCF.

The data in Fig. 2 exhibited along the time axis are segmented with respect to the roundtrip time and then the buildup dynamics of solitons can be depicted with the roundtrip time and the roundtrip number simultaneously. Then the real-time evolution dynamics of CS, SP and DS are shown in Figs. 3, 4 and 5, respectively.

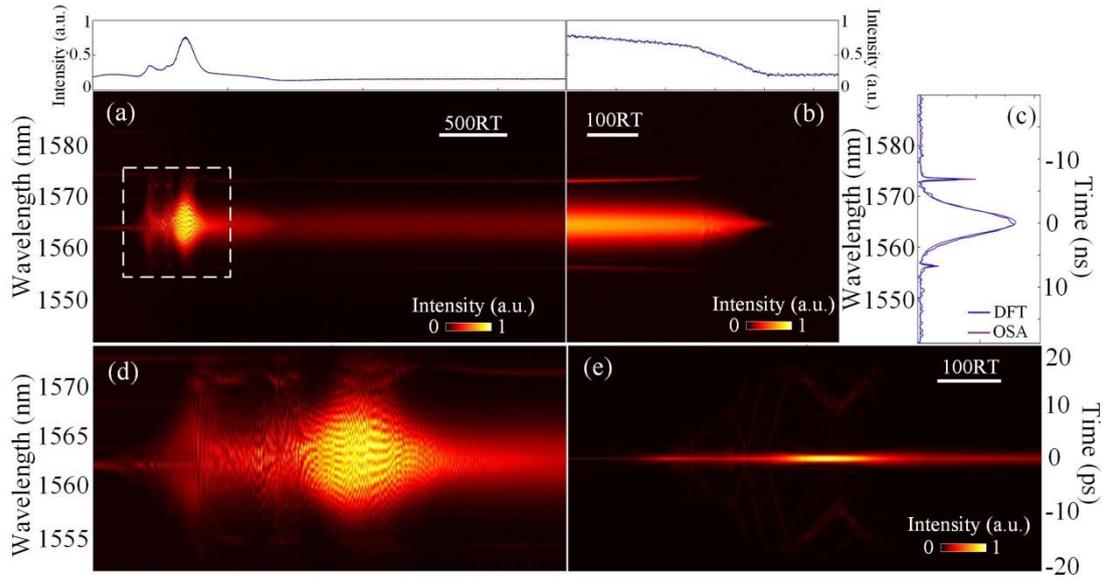

Fig. 3. Starting and shutting dynamics for CS fiber laser. (a) Real-time spectral evolution dynamics during the formation of CS. The above picture shows the corresponding energy of every roundtrip in (a). (b) Real-time spectral evolution dynamics during the shutting for CS. The above picture shows the corresponding energy of every roundtrip in (b). (c) Optical spectra of soliton measured by OSA and DFT technique. (d) Close-up of the data in the square from (a), revealing the interference pattern for the beating dynamics and the complex multi-pulse evolution. (e) The Fourier transform of each single-shot spectrum corresponds to the field autocorrelation of (d).

The experimental data in Fig. 3 are from Fig. 2(a), which shows the starting and shutting dynamics for CS in the SWNT-mode-locked fiber laser. As that in the previous results [2][21], beating dynamics is firstly formed with the spectrum broadening, and then laser experience a complex multi-pulse evolution as shown in Figs. 3(d) and (e). Especially, the complex multi-pulse spectrum can grow continuously in an extremely short time, which corresponds to the high lasing spike in Fig. 1(a). Finally, the complex multi-pulse state collapses and forms the stable single-soliton operation. The complex pulse and energy evolution reflects the variation of gain that is related to the population inversion in EDF [31]. The longer EDF in this work could lead to the strong gain oscillation. Strong amplification of pulses lead to the transient high peak power inducing the strong self-phase modulation and the abrupt spectral broadening. And CS cannot maintain the single-pulse operation state. From the shutting process in Fig. 3(b), the decaying process of soliton can be resolved clearly. As the pulse energy decreases, the spectral width and intensity become smaller. The sidebands depart from the central wavelength gradually and disappear at a certain position. Figure 3(c) shows the spectra of CS measured by OSA and single-shot spectrum measured via DFT technique. The real-time single-shot spectrum agrees quite well with the time-averaged optical spectrum, highlighting the mapping relationship linked by the dispersion [19]. The experimental observations show that the spectra have obvious Kelly sidebands, which are the typical characteristics of solitons [32].

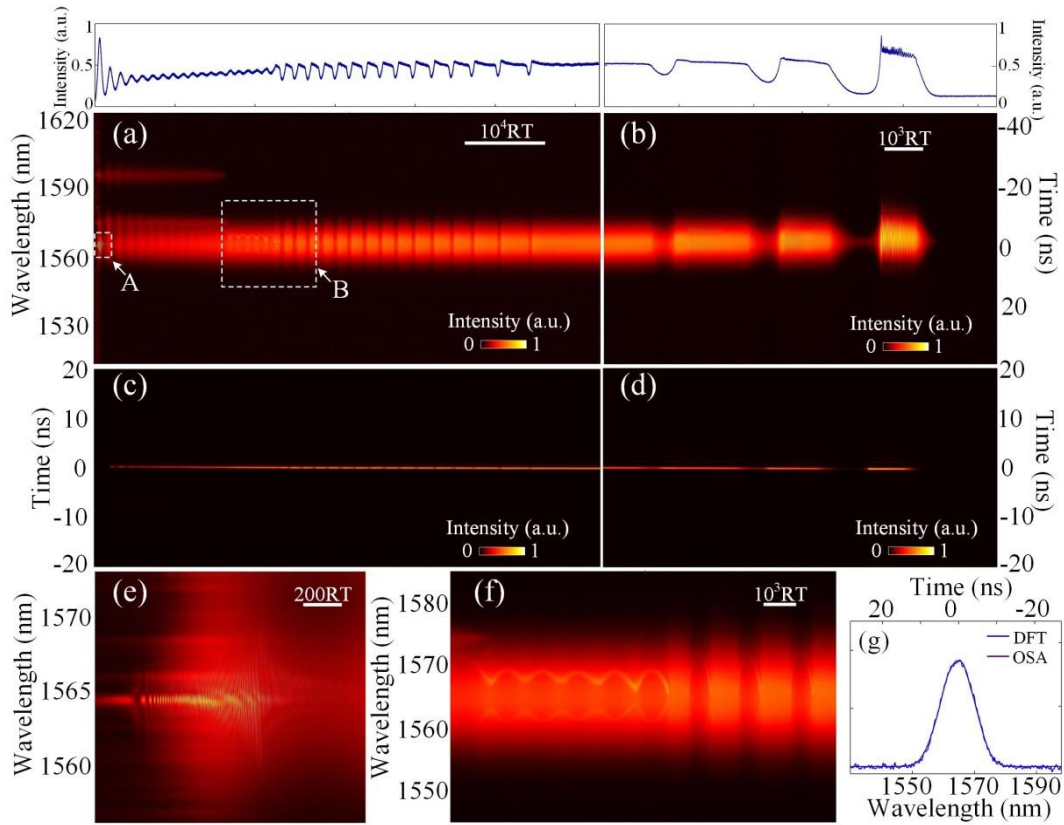

Fig. 4. Starting and shutting dynamics for SP fiber laser. (a) Real-time spectral evolution dynamics during the formation of SP. The above picture shows the corresponding energy of every roundtrip in (a). (b) Real-time spectral evolution dynamics during the shutting process for SP. The above picture shows the corresponding energy of every roundtrip in (b). Experimental real-time observation without DFT for (c) starting and (d) shutting processes, corresponding to (a) and (b), respectively. (e) Close-ups of the data from A region in (a), revealing the interference pattern for the beating dynamics. (f) Interaction of SPs and picosecond pulses. The data are from B region in (a). (g) Optical spectra of SP measured by OSA and DFT technique.

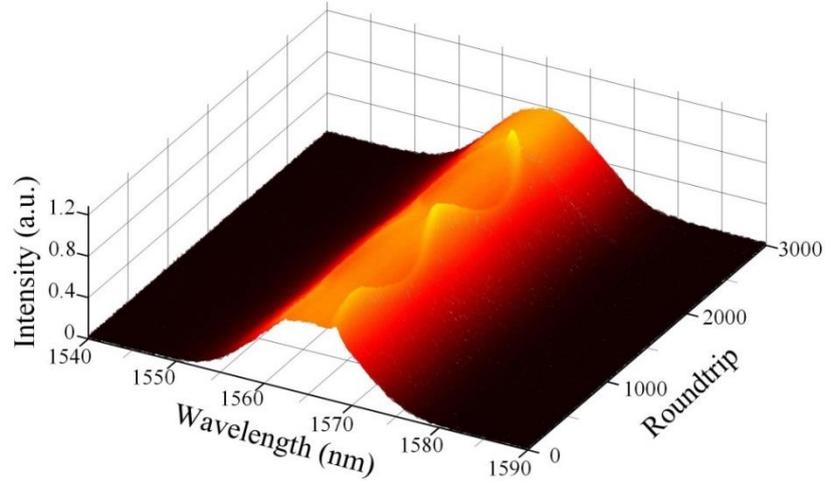

Fig. 5. Interaction of SPs and picosecond pulses. Data are from Fig. 4(f).

Figure 4 shows the starting and shutting dynamics for SP, corresponding to Fig. 2(b). Beating dynamics also exist in the build-up process of SP shown in Figs. 4(a) and (e). After the beating dynamics, the spectral bandwidth and energy increase gradually within ~200 roundtrips, which is different from CS (Fig. 3(a)). And the evolving SP is accompanied by two subordinate pulses. From Figs. 4(a) and (c), we can observe that obvious periodical modulation is displayed along the roundtrip, even for the remained subordinate pulses. And the energy modulation after the formation of SP is consistent with that of the relaxation oscillation in the above figure of Fig. 4(a). It further confirm the inference that the modulation stems from the gain variation induced by relaxation oscillation. Additionally, there is an interaction process between SPs and picosecond pulse during the periodical modulation stage, as shown in Fig. 4(f). From the zoom-in typical results in Fig. 5. we can see the wavelengths of two picosecond pulses shift with the periodical energy modulation. When the energy becomes larger (smaller), the wavelengths of the picosecond pulses move toward (backward) the central wavelength of SP. The spectral and temporal shutting dynamics are shown in Fig. 4(b) and (d), respectively. Corresponding to the energy fluctuation in temporal domain in Fig. 2(b) and 4(d), the spectral width and intensity change as shown in Fig. 4(b). Note that SPs would operate in an unstable manner during the shutting process due to the drastic change of population inversion. Figure 4(g) shows the spectra of SP measured by OSA and DFT technique, and they show the similar profile with a smooth Gaussian-shape. It is the typical characteristics of SPs [7][11].

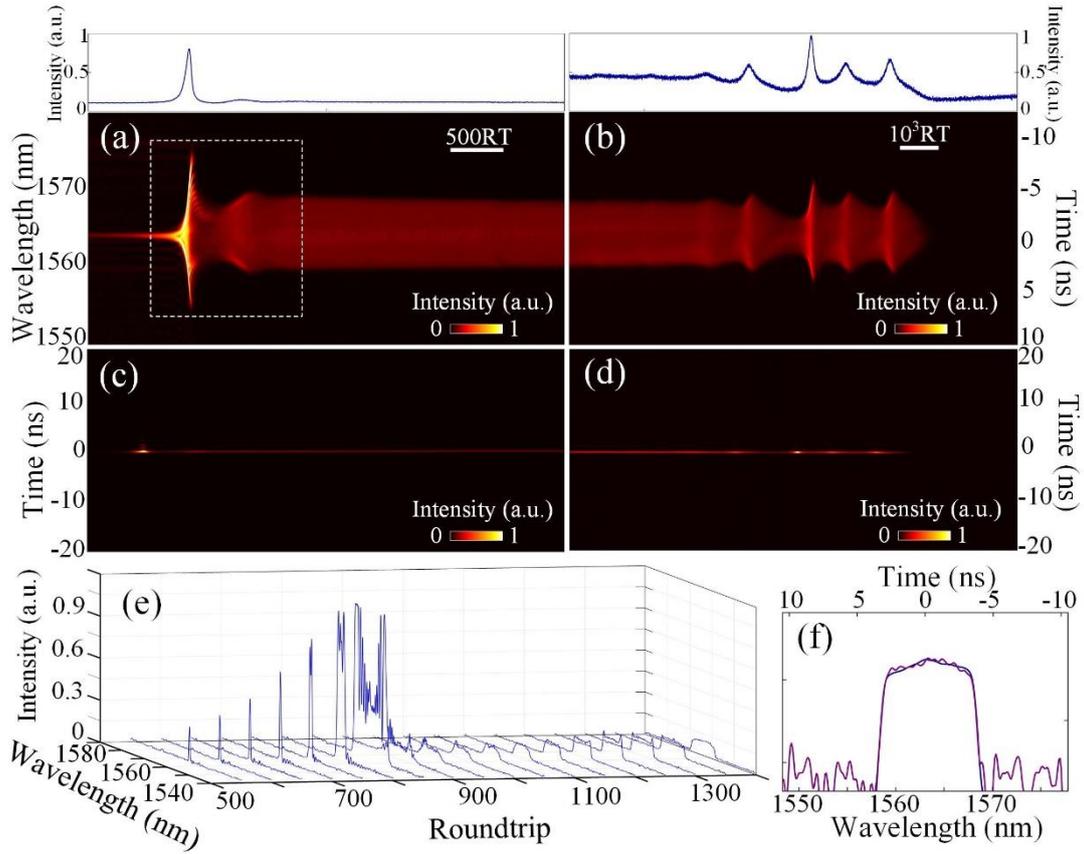

Fig. 6. Starting and shutting dynamics for DS fiber laser. (a) Real-time spectral evolution dynamics during the starting of DS. The above picture shows the corresponding energy of every roundtrip in (a). (b) Real-time spectral evolution dynamics during the shutting for DS. The above picture shows the corresponding energy of every roundtrip in (b). Experimental real-time observation without DFT for (c) starting and (d) shutting processes, corresponding to (a) and (b), respectively. (e) Spectral evolution for the formation of DS. The data are extracted from the region within the square in (a). (f) Optical spectra of DS measured by an OSA and DFT technique.

Figure 6 shows the starting and shutting dynamics for DS, corresponding to Fig. 2(c). The spectra broaden in a short time along with the rapid increase of intra-cavity energy. Some interference fringes can be observed at this stage in Fig. 6(a), which is distinct from the beating dynamics in Figs. 3 and 4. A series of typical spectra are extracted from the Fig. 6(a) and depicted in Fig. 6(e) along with the roundtrip. The single-shot spectra at both edges have the evident fringes that could be generated when the pulse energy of DS is large enough [7]. Then the spectral width broadens and compresses with the energy, and finally stable DS forms. During the shutting process, the spectral width of DS shows oscillation due to the variation of the energy. In fact, the remained population inversion can still provide gain within the relaxation time of doped ions. But it is difficult to drive the stable operation. By comparison with SP, DS can experience more roundtrips because of the shorter time for every roundtrip.

**Conclusions**

CS, SP and DS are achieved in the SWNT-mode-locked fiber laser under the different dispersion conditions. By utilizing pump power as the triggering signal, the starting and shutting processes of CS, SP and DS are measured via TS-DFT technique. The relaxation oscillation can be observed in the build-up processes of CS, SP and DS benefiting from the used SWNT SA. However, the transition stage exhibits the distinct evolution pathway from the relaxation oscillation to the stable mode-locking operation. After the beating dynamics, the spectrum broadens abruptly with the complex interaction of multiple pulses in CS mode-locked fiber laser. SP can form directly via beating dynamics, but is induced a long instability by the relaxation oscillation. However, beating dynamics cannot be observed in the starting process of DS. For the shutting process, CS, SP and DS also experience the different dynamics and energy fluctuation. CSs vanish quickly, while SPs and DSs experience a long Q-switched fluctuation before the extinction. These findings provide new perspectives into the ultrafast transient dynamics and bring real-time insights into laser design and applications.